\documentclass[12pt]{article}

\usepackage{sbc-template}

\usepackage{graphicx,url}
\usepackage[colorlinks = true,
            linkcolor = blue,
            urlcolor  = blue,
            citecolor = blue,
            anchorcolor = blue]{hyperref}

\usepackage[utf8]{inputenc}  
\usepackage{doi}
     
\title{Specification for the Siril HEALpixel Catalog Format\\
1.0.0}

\author{Knagg-Baugh Adrian J. E. \and Cass Ian \and Melis Cécile\and Richard Cyril\inst{1}}

\address{Laboratoire Interdisciplinaire Carnot de Bourgogne\\
UMR 6303 CNRS, Universit\'{e} Bourgogne Europe,\\
9 Av. A. Savary, BP47870, F-21078 Dijon Cedex, France
}

\begin{document} 

\maketitle

\begin{abstract}
This document specifies the structure of the Siril Catalog Format, designed to support efficient storage and querying of nested HEALpixel based astronomical catalogs such as Gaia DR3. The format includes a fixed length header, an index for rapid look-up, and structured data records optimized for space and speed. The document also outlines recommended search strategies for utilizing the format effectively.
\end{abstract}
     
\section{Introduction}

\subsection*{Acknowledgment}
This work has made use of data from the European Space Agency (ESA) mission Gaia\footnote{\url{https://www.cosmos.esa.int/gaia}}, processed by the Gaia Data Processing and Analysis Consortium\footnote{\url{https://www.cosmos.esa.int/web/gaia/dpac/consortium}} (DPAC). Funding for the DPAC has been provided by national institutions, in particular the institutions participating in the Gaia Multilateral Agreement~\cite{Gaia_2016, Gaia_2023}.

\subsection*{Purpose}
This document defines the Siril Catalog Format, enabling users to store and access large astronomical catalogs efficiently. The format is designed for high-performance use cases, such as astrometric and photometric data extraction and analysis.

\subsection*{Background}
The format is intended to work seamlessly with Siril~\cite{siril}, a free and open-source tool for astronomical imaging and processing. It primarily accommodates data from Gaia DR3~\cite{gaia} and subsequent data releases, allowing flexible integration with nested HEALpixel~\cite{gorski1999,gorski1999healpix} based searches for spatial queries.

\subsection*{Scope}
This specification covers the format’s structure, data types, and intended use. It does not address broader interoperability with other formats or general data processing techniques.

\subsection*{Version History}
\begin{itemize}
    \item \textbf{1.0.0}: this is the initial version of the catalog specification. When the specification is updated, subsequent versions of this document will summarize the key changes throughout the version history.
\end{itemize}

\section{Data Format Overview}
\subsection*{Design Principles}
\begin{itemize}
    \item Efficient for storage and search.
    \item Structured for compatibility with Gaia data releases.
    \item Extensible to allow future updates without breaking compatibility.
    \item Supports customizing catalog content to match the use case. For example, an annotations catalog may opt for encyclopaedic coverage down to a specified limiting magnitude whereas a plate solving catalog may target a set number of sources per index HEALpixel in order to achieve even coverage for plate solving without including an excessive number of sources in densely populated sky regions.
\end{itemize}

\subsection*{Typical Use Cases}
\begin{itemize}
    \item Astrometric queries within a predefined region of the sky. 
    \item Spectrophotometric data extraction for subsets of celestial sources. 
\end{itemize}

\subsection*{Key Features}
    \begin{itemize}
        \item Fixed-length headers for rapid identification.
        \item Catalog type defines the content of each record. This is specified within the catalog header using an enum, but the definition of the record content is defined externally to the catalog file.
        \item Index structure for quick healpixel-based data lookup.
        \item Support for dividing large catalogs into chunks based on a lower HEALpixel level.
        \item Compact, sorted data records to minimize storage and access overhead.
    \end{itemize}

\section{File Structure}
\subsection*{HEADER}
The header is a fixed 128 \verb|bytes| as follows:
    \begin{itemize}
    \item {48 \verb|bytes|}: Catalog title (e.g., \verb|"Siril Gaia DR3 astrometric extract"|).
    \item {1 \verb|byte|}: Gaia version designator as an integer enum with values defined as follows:
        \begin{itemize}
            \item DR1 = 0
            \item DR2 = 1
            \item eDR3 = 2
            \item DR3 = 3
            \item DR4 = 4
            \item DR5 = 5
        \end{itemize}
    \item {1 \verb|uint8_t|} HEALpixel indexing level, denoted as N. For Gaia catalogs, the indexing level is between 1 and 12. The initial Siril catalogs will adopt an indexing level of 8.
    \item {1 \verb|uint8_t|}: Catalog type designator as an integer enum with values defined as follows:
    \begin{itemize}
        \item Astrometric extract = 1
        \item Photometric extract with \verb|xp_sampled| data = 2
        \item Photometric extract with \verb|xp_continuous| data = 3
        \item Values of 4+ are available for future use. It is recommended that applications other than Siril use local values $>=$ 128 to avoid any conflict.
    \end{itemize}
    \item {1 \verb|uint8_t|}: Indicates whether the catalog is chunked or not. Must be zero for monolithic catalogs and non-zero for chunked catalogs.
    \item {1 \verb|uint8_t|}: For chunked catalogs, the level at which the catalog is chunked (e.g. if the catalog is chunked at HEALpix level 1 and therefore split across 48 files, this byte would be 1). Set to zero for monolithic catalogs.
    \item {1 \verb|uint32_t|}: For chunked catalogs, the chunk-level HEALpixel covered by this specific file (e.g. for a catalog chunked at HEALpix level 1 this might be any number between 0 and 47). Set to zero for monolithic catalogs.
    \item {1 \verb|uint32_t|}: For chunked catalogs, the first HEALpixel number covered by this file at the indexing HEALpixel level. Set to zero for monolithic catalogs.
    \item {1 \verb|uint32_t|}: For chunked catalogs, the last HEALpixel number covered by this file at the indexing HEALpixel level. Set to zero for monolithic catalogs.
    \item {63 \verb|bytes|}: Reserved for future use.
    \end{itemize}
    
\subsection*{INDEX}
\begin{itemize}
    \item Size: \verb|N_HEALPIXELS_AT_LEVEL_N| $\times$ \verb|sizeof(uint32_t)| bytes. For example, at level 8 there are 28 × 28 × 12 = 786,432 HEALpixels, so the index size is 786432 × 4 = 3,145,728 bytes.
    \item Content: For each level N HEALpixel in numerical order, store a \verb|uint32_t| representing the cumulative number of sources up to and including that HEALpixel.
\end{itemize}

\subsection*{DATA}
The different catalog types can hold different and arbitrary data structures. Each source must occupy a fixed number of bytes. The data structures of the initial two Siril catalogs (the astrometric extract~\cite{siril_astro_cat} and the sampled spectrophotometric extract~\cite{siril_spcc_cat}) are as follows:
\paragraph{Astrometric Extract}
    \begin{itemize}
        \item \verb|int32_t RA_scaled|: Scaled Right Ascension in degrees. A scaling factor of $360\ / (2^{31}-1)$ is applied.
        \item \verb|int32_t Dec_scaled|: Scaled Declination in degrees. A scaling factor of $360\ / (2^{31}-1)$ is applied.
        \item \verb|int16_t dRA|: Proper motion in Right Ascension (mas/yr), rounded to the nearest integer.
        \item \verb|int16_t dDec|: Proper motion in Declination (mas/yr), rounded to the nearest integer.
        \item \verb|uint16_t Teff|: Effective temperature in Kelvin (or 0 if not available).
        \item \verb|int16_t G_mean_mag|: Scaled mean magnitude in the Gaia G band. A scaling factor of $1 \times 10^3$ is applied.
        \item Total size per record: 16 \verb|bytes|.
    \end{itemize}
    The rationale for this structure is to keep as much commonality as possible with existing code that processes KStars binary catalogs~\cite{kstars}, however whereas the KStars catalogs use multipliers of $1 \times 10^6$ for RA and $1 \times 10^5$ for Dec, the Siril catalogs use a common multiplier of $360\ / (2^{31}-1)$ for both quantities. This provides a constant angular resolution of $\approx$ 0.60 mas for all angles between 0 and $360^\circ$. Note that the structure does not store the \verb|gaia_source source_id| field. This is because for the intended purpose of the catalog it is not required: HEALpixels are indexed at the front of the catalog, but we don't care what order the sources are presented in, and 8 bytes per source is a significant addition to the 16-byte record size.
    
\paragraph{Spectrophotometric Extract with \texttt{xp\_sampled} data}
    \begin{itemize}
        \item \verb|int32_t RA_scaled|: Scaled Right Ascension in degrees. A scaling factor of $360\ / (2^{31}-1)$ is applied.
        \item \verb|int32_t Dec_scaled|: Scaled Declination in degrees. A scaling factor of $360\ / (2^{31}-1)$ is applied.
        \item \verb|int16_t dRA|: Proper motion in Right Ascension (mas/yr), rounded to the nearest integer.
        \item \verb|int16_t dDec|: Proper motion in Declination (mas/yr), rounded to the nearest integer.
        \item \verb|int16_t G_mean_mag|: Scaled mean magnitude in the Gaia G band. A scaling factor of $1 \times 10^3$ is applied.
        \item \verb|int8_t fexpo|: Exponent for use with the following flux coefficients.
        \item \verb|uint16_t flux[343]|: \verb|float16| flux data at each sampling wavelength (from 336nm to 1020nm in 2nm intervals).
        \item Total size per record: 701 \verb|bytes|.
    \end{itemize}
    This structure is based on the preceding astrometric structure, however Teff is omitted and a representation of the Gaia DR3 \verb|xp_sampled| data is included. To optimize the space taken up by the catalog, the \verb|xp_sampled| data is represented by a half precision float (\verb|float16|). Since the float16 format cannot handle the exponent values required by the \verb|xp_sampled data|, an \verb|int8_t| provides an overall exponent to be applied to all 343 values. Conversion to \verb|float16| from \verb|float32| does lose some numerical precision, however this is justified by the magnitude of the errors associated with \verb|xp_sampled| data, and testing of the spectrophotometric color calibration process in Siril has shown no difference between the local dataset using \verb|float16| and using the original Gaia archive dataset at \verb|float32| precision.

\paragraph{Photometric Extract with \texttt{xp\_continuous} data}
\begin{itemize}
    \item To be defined in a future version of this specification.
\end{itemize}

\subsection*{Sorting Requirement}
Records belonging to individual HEALpixels must be grouped together and groups must occur in the catalog in HEALpixel order, with numbers of records represented as a cumulative count in the index section. Within a given indexed HEALpixel the order of sources does not matter.

\section{Data Types and Encoding}
\subsection*{Supported Data Types}
    \begin{itemize}
        \item \verb|int8_t|: Signed 8-bit integer.
        \item \verb|uint8_t|: Unsigned 8-bit integer.
        \item \verb|int16_t|: Signed 16-bit integer.
        \item \verb|uint16_t|: Unsigned 16-bit integer.
        \item \verb|int32_t|: Signed 32-bit integer.
        \item \verb|uint32_t|: Unsigned 32-bit integer.
        \item \verb|int64_t|: Signed 64-bit integer.
        \item \verb|uint64_t|: Unsigned 64-bit integer.
        \item \verb|float16|: Half precision 16-bit floating point number (IEEE 754~\cite{IEEE754-2019}).
        \item \verb|float32|: Single precision 32-bit floating point number (IEEE 754).
        \item \verb|float64|: Double precision 64-bit floating point number (IEEE 754).
    \end{itemize}
    Note that not all of these data types are used in the initial Siril catalog releases.

\subsection*{Encoding Details}
    \begin{itemize}
        \item All data are stored in little-endian format.
        \item RA and Dec are scaled to improve precision during storage and require appropriate scaling factors during decoding.
    \end{itemize}

\section{Metadata Specifications}
The following metadata fields are mandatory in the header (see Section 3):
    \begin{itemize}
        \item Catalog Title: Human-readable identifier, ASCII encoded.
        \item Gaia Data Release: Enum specifying the Gaia data release.
        \item Healpix Indexing Level (N): Integer defining the spatial resolution.
        \item Catalog Type: Indicator of astrometric or photometric data.
        \item Reserved Bytes: Must be initialized to zero if not required (they are required in chunked catalogs).
    \end{itemize}

\section{Examples}
\subsection*{Example Header}
\begin{verbatim}
"Siril Gaia DR3 astrometric extract" (48 bytes)
0x03 (Gaia DR3)
0x08 (Healpix Level 8)
0x01 (Astrometric catalogue)
77 bytes of zero padding
\end{verbatim}

\subsection*{Example Index}
For Healpix Level 8 with the first five pixels containing 10, 20, 0, 15, and 30 sources respectively:
\begin{verbatim}
[0x0000000A, 0x0000001E, 0x0000001E, 0x00000033, 0x00000051, ...]
\end{verbatim}

\subsection*{Example Astrometric Extract Data Record}
\begin{verbatim}
RA_scaled: 123456789
Dec_scaled: -987654321
dRA: 12
dDec: -18
Teff: 6000
G_mean_mag: -20
\end{verbatim}

\section{Search Strategy}
Efficient search strategies make use of the property of the nested HEALpixel scheme used by the Gaia catalog, which means that the HEALpix numbers of HEALpixels in a similar part of the sky will be close together.

\subsection*{Monolithic Catalogs}
    \begin{itemize}
        \item Conduct a conesearch that returns a list of HEALpixels (at the same HEALpixel level as the catalog is indexed) touching the search area.
        \item Reduce the list of individual HEALpixels to a list of ranges of consecutive HEALpixels.
        \item For each range of consecutive HEALpixels, look up the start and end indices and calculate the number n of records to be read.
        \item Seek to the start index, accounting for the size of the header and the index.
        \item Read n records directly into memory cast as an array of structs representing the catalogued data.
        \item Perform any necessary filtering (e.g. by scaled magnitude or by distance from the center of the search)
        \item Scale the data values of the matching sources accordingly.
    \end{itemize}

\subsection*{Chunked Catalogs}
    \begin{itemize}
        \item Conduct a conesearch that returns a list of HEALpixels (at the same HEALpixel level as the catalog is indexed) touching the search area.
        \item Scan the list of HEALpixels and produce a list of chunk-level HEALpixels that contain matches. Sort the conesearch results into lists corresponding to each chunk-level HEALpixel.
        \item For each chunk, reduce the list of individual HEALpixels to a list of consecutive ranges and proceed in accordance with the monolithic catalog search strategy above, using the correct file for the catalogue chunk in question.
        \item Concatenate the results from each chunk into an overall set of results.
    \end{itemize}

\subsection*{Online / Hybrid Online + Local Use}
Catalogs implementing this format may be stored online and read efficiently using HTTP RANGE requests to read the data from specified ranges in the file.
A hybrid strategy could be implemented where a catalog is hosted online and as the user makes queries online a sparse file is built up locally providing a cache that can be checked to avoid slower online queries for subsequent use. This strategy is proposed for implementation during the Siril 1.5 development cycle.

\section{Compatibility and Extensions}
\subsection*{Backward Compatibility}
Future updates will ensure that:
    \begin{itemize}
        \item Unused header bytes retain compatibility.
        \item Extensions are appended without altering existing structure.
    \end{itemize}

\subsection*{Custom Fields}
In order to create a catalog with different fields, the catalog should be assigned a different catalog type specifier in the header. The required fields must be written to the catalog as fixed size packed structs, but the details of what a catalog entry may contain are left to the implementer.

\section{Best Practices}
Siril catalog files should follow the convention:

\verb|siril_cat<chunk_level>_healpix<level>_<type>_<N>.dat|.

Here, \verb|<chunk_level>| and \verb|<N>| may be omitted for monolithic catalogs; \verb|<type>| can currently be "\verb|astro|", "\verb|xpsamp|" or "\verb|xpcts|". Other \verb|<type>| strings may be locally defined for local catalog types.

\subsection*{Examples}
    \begin{itemize}
        \item \verb|siril_cat_healpix8_astro.dat| -- a monolithic astrometry catalog indexed at HEALpixel level 8.
        \item \verb|siril_cat2_healpix8_xpsamp_43.dat| -- chunk 43 of a photometric catalog containing \verb|xp_sampled| data indexed at HEALpix level 8 and chunked at HEALpix level 1 to give 48 catalog chunk files.
    \end{itemize}

\section{Data Integrity}
The initial Siril catalogs were compiled by importing the \verb|gaia_source| table (and, in the case of the \verb|xp_sampled| catalog, the \verb|xp_sampled| datalink product) into a local SQL database and using python to extract the top 127 records sorted by \verb|G_mean_mag| from each level 8 HEALpixel. The xp\_sampled catalog is split into chunks each covering one level 1 HEALpixel. Automated validation scripts were generated using pytest and a sample of the records were cross-checked against the original SQL database~\cite{Cass2025}.

Catalogs should be published together with a checksum that can be used to verify their integrity. The initial releases of the Siril astrometric and \verb|xp_sampled| catalogs are published with sha256sums for both the compressed and uncompressed files.

\section{Implementation Guidance}
\subsection*{Libraries/Code}
    \begin{itemize}
        \item Use of these catalogs requires a library that support efficient HEALpixel-based searching. Within Siril a subset of \verb|libhealpix_cxx|~\cite{libhealpix} is used.
        \item The reference implementation can be found at \url{https://gitlab.com/free-astro/siril/-/blob/master/src/io/healpix/healpix.cpp}
    \end{itemize}

\section{Licensing and Citation}
    \begin{itemize}
        \item License: CC-BY-4.0 (Creative Commons Attribution).
        \item Citation: Cite this document using its DOI and version number:
        \item The DOI 10.5281/zenodo.14697485 will always refer to the latest version of the document.
        \item The DOI 10.5281/zenodo.14697486 refers to this version (1.0.0) of the document.
    \end{itemize}

\section{Glossary}
    \begin{itemize}
        \item HEALpix: Hierarchical Equal Area IsoLatitude Pixelization.
        \item RA/Dec: Right Ascension and Declination, celestial co-ordinates.
        \item DR: Data Release (of Gaia data).
    \end{itemize}

\section{References}
\renewcommand{\refname}{}
    \bibliographystyle{plainurl}
    \bibliography{biblio}

\end{document}